\documentclass[aip,jcp,amsmath,amsymb,reprint]{revtex4-1}
\usepackage{graphicx}
\usepackage{bm}

\begin{document}
	\title{Inverse Design of Multicomponent Assemblies}
	\author{William D. Pi\~{n}eros} 
	\affiliation{Department of Chemistry, University of Texas at Austin, Austin, TX 78712}
    \author{Beth A. Lindquist}
	\affiliation{McKetta Department of Chemical Engineering, University of Texas at Austin, Austin, TX 78712}
	\author{Ryan B. Jadrich}
	\affiliation{McKetta Department of Chemical Engineering, University of Texas at Austin, Austin, TX 78712}
	\author{Thomas M. Truskett}
	\affiliation{McKetta Department of Chemical Engineering, University of Texas at Austin, Austin, TX 78712}
	\affiliation{Department of Physics, University of Texas at Austin, Austin, TX 78712}
	\date{\today}
	\begin{abstract}
Inverse design can be a useful strategy for discovering interactions that drive particles to spontaneously self-assemble into a desired structure. Here, we extend an inverse design methodology--relative entropy optimization--to determine isotropic interactions that promote assembly of targeted multicomponent phases, and we apply this extension to design interactions for a variety of binary crystals ranging from compact triangular and square architectures to highly open structures with dodecagonal and octadecagonal motifs. We compare the resulting optimized (self and cross) interactions for the binary assemblies to those obtained from optimization of analogous single-component systems. This comparison reveals that self interactions act as a `primer' to position particles at approximately correct coordination shell distances, while cross interactions act as the `binder' that refines and locks the system into the desired configuration. For simpler binary targets, it is possible to successfully design self-assembling systems while restricting one of these interaction types to be a hard-core-like %(Weeks-Chandler-Andersen) 
potential. However, optimization of both self and cross interaction types appears necessary to design for assembly of more complex or open structures. 
	\end{abstract}

	\maketitle %\maketitle must follow title, authors, abstract and \pacs

%--Introduction
\section{Introduction} 
	The synthesis and fabrication of materials with specific sub-mesoscale features or architectures is a desirable but challenging design goal. While some top-down techniques allow material patterning at these scales (e.g., lithography), they are often prohibitively slow and expensive processes for large-scale industrial applications. \cite{PhotonicMatsDesign,PhotonicMatsDesign2,NanoFabReview} Self-assembly of colloidal systems, on the other hand, presents one promising bottom-up alternative whereby particles are designed to organize into specific configurations determined by their effective interparticle interactions.\cite{SelfAssemblyPerspective,SelfAssemblyReviewColloids,SelfAssemblyReviewColloids2,SelfAssemblyForcesReview}  This approach, though still in its infancy, is attractive due to the possibility to systematically tune the effective interparticle interactions\cite{ColloidInteractionsRev,ColloidInteractionTuning_1,ColloidInteractionTuning_2}--including those determined by particle shape\cite{SelfAssemblyPolyhedraParticles,GeometricParticlesAssembly,id_jamming_1,id_jamming_2} or ``patchiness''\cite{SelfAssemblyPatchyParticles1,JanusParticlesSelfAssemblyRev,SelfAssembly_Patchy3D_Open_Structures,patchy_review1,kagome,capsid}--to drive self-assembly. It is also possible to expand the available design space for assembly by employing multicomponent systems (e.g., DNA-grafted nanoparticles\cite{DNA_nanoparticle_crystal,gold_binary_DNA_directed_assembly,DNA_binary_nanoparticle_control}, binary systems of charged colloids\cite{charged_binary_nanoparticle,charged_binary_colloid,protein_cage_charged_binary_colloid_assembly} and  others\cite{nanorods_sphere_template_assembly,2step_multicomponent_clusters_ordered_superstructures,binary_nanocomposite_magnets,binary_copolymer_assembly}) to augment the diversity of possible interactions and structures. 
	
	Given the large potential parameter space available, systematic design strategies that allow one to explore how the properties of constituent particles relate to the resulting self-assembled structures are needed. To this end, one can consider \emph{forward} design methods whereby the phase behavior of a system with known interactions is mapped by systematic variation of a few characteristic parameters. Alternatively, \emph{inverse} design methods can be employed, where a desired particle assembly is explicitly targeted, and the parameters necessary to achieve it are found by way of solving a constrained optimization problem.\cite{InvDesignPerspective,InvDesignTechRev,InvDesignGeneral,nucleation_id,patchy_screening,template_directed,property_id,swarm_id,id_jamming_1,id_jamming_2,evolution_polymer}   

	A classic inverse design problem is the optimization of decision variables $\boldsymbol{\theta}$ for a system of identical particles interacting via a given isotropic pairwise interaction potential, $\phi(r;\boldsymbol{\theta})$, to stabilize a desired structure. Here, $r$ is the distance between particle centers and $\boldsymbol{\theta}$ are the parameters required for the pair potential. By minimizing either the ground state energy of the ideal configuration (relative to competing structures) or the free energy of an associated configurational ensemble at a higher temperature (relative to competing phases), researchers have successfully found isotropic interactions that stabilize a wide variety of structures and phases including, for example, two-dimensional honeycomb\cite{RT_HoneyDoubleWell,MT_SquareHoneyConvexFull,AvniDimTransfer} and kagome\cite{InvDesignKagome,InvDesignKagomeFunctionalMethod,ZT_MuOptKagomeAsymLats} lattice assemblies as well as three-dimensional simple cubic\cite{Avni3DLattices} and diamond\cite{Avni3DLattices,MT_DiamondConvex,InvDesignKagomeDiamond} crystals. Isotropic pair potentials that stabilize more exotic phases have also been discovered via recently introduced inverse design strategies\cite{SquLat_dmu_opt,RelEntropy_2D_structures,RelEntropy_general}. An analytical reformulation\cite{SquLat_dmu_opt} of the ground-state optimization problem\cite{Avni3DLattices} allowed for the efficient design of potentials that stabilize two-dimensional snub-square\cite{KagSnub_gams_opt} and truncated hexagonal \cite{Truncs_gs_opt} lattices. Furthermore, a strategy adapted from the bio-molecular coarse graining community\cite{IBI1,IBI2,IBI3,test_of_ID_schemes,voth_ID} called relative entropy (RE) minimization\cite{RE_ID,general_RE}--which enables ``on-the-fly'' optimization of the potential parameters directly within a molecular simulation--led to the discovery of isotropic interactions that promote self-assembly of a variety of two and three dimensional crystals and quasicrystals \cite{RelEntropy_2D_structures,RelEntropy_general,RelEntropy_FKphases} as well as clustered fluids and porous mesophases.\cite{inv_design_clusters,inv_design_porous_mesophases,RelEntropy_general} 
	
	While much of the earlier inverse design work focused on the wealth of structural possibilities for single-component systems, similar demonstrations in multicomponent systems are surprisingly lacking. Multicomponent systems are of particular interest given the expanded design space afforded to them through their additional self and cross component interactions (i.e., additional degrees of freedom). Moreover, the ability to partition a desirable target structure into a larger number of components suggests it may help simplify the particle interactions required for a given structure relative to those of an equivalent single-component system. Indeed, multicomponent systems are known to stabilize entirely new and exotic structures which are inaccessible to single-component systems. For instance, it is known that moving from single-component to binary colloidal systems significantly diversifies the list of crystal structures and motifs observed in experiment.\cite{binary_nanoparticle_diversity,binary_nanoparticle_diversity2,DNA_binary_nanoparticle_control} Similarly, recent theoretical and computational efforts demonstrate that binary mixtures can produce a wide variety of novel structures, some of which exhibit highly specific local orderings. \cite{2D_binary_theoretical_assembly,theoretical_binary_phase_diagram,3D_binary_hardcore_vdw_simulations,Binary_phase_diagram_soft_potentials,ArchStructures_BinaryMix}

	Clearly, multi-component systems hold signficant promise as tailorable building blocks for novel structural architectures, but a design strategy that allows for rational design of underlying interactions remains to be demonstrated. Here, we adopt one such strategy, and extend the RE optimization method for inverse design of single-component materials interacting via isotropic interparticle potentials to multicomponent systems. By independently optimizing the self and cross interactions for species in a binary two-dimensional mixture, we demonstrate that this methodology can successfully discover pair potentials that readily self-assemble a variety of simple (intercalated square and triangular) lattices as well as more challenging target phases (stripes and highly open structures), many of which had not yet, to our knowledge, been observed to self-assemble for such systems. For select cases, we further compare the optimized binary interactions to those obtained from optimizations for the target structures assuming single-component systems. These comparisons help to understand the trade-offs associated with using binary systems for self assembly as well as the roles of self versus cross interactions in stabilizing various target lattices.

	The remainder of this paper is organized as follows. We elaborate on the extension of the RE optimization strategy to multicomponent systems and explain how our binary crystal target structures are chosen in Sect.~\ref{sec:Methods}. We then show the optimized pair potentials and assemblies for all nine successfully assembled targets and discuss for select cases how binary interactions compare to those of analogous single-component interaction features at a global, and local, sub-lattice structure level in Sect.~\ref{sec:Results}. We conclude in Sect.~\ref{sec:Conclusion} by summarizing the similarities and differences between the binary and the single component interaction potentials, focusing specifically on how these observed features enable formation of the targeted structures. 

%--Methods
\section{Methods} 
\label{sec:Methods}

\subsection{Relative Entropy Optimization}
The RE course graining approach is a probabilistic optimization method that addresses an inverse design problem for self-assembly by systematically tuning the interparticle interaction potential $U(r|\boldsymbol\theta)$ in a system of particles via parameters $\boldsymbol\theta \equiv \{\theta_1,...\theta_m\}$ to maximize the likelihood of forming a desired structure. One advantage of the approach is that it can perform ``on-the-fly'' optimization of particle interactions during the course of a simulation in order to promote thermodynamic stability (and, with a judicious choice for the simulation protocol,\cite{RelEntropy_2D_structures} kinetic accessibility) of the assembled target phase. Although RE optimization was originally applied to single-component systems,\cite{RelEntropy_2D_structures,RelEntropy_general} it can be readily extended to multicomponent materials. For a two-dimensional binary system with components $A$ and $B$, we partition the energy and parameters in terms of self and cross interactions as $u(r|\boldsymbol\theta^{(k,k')})$ where $k,k'=A\;\text{or}\;B$. Parameters $\boldsymbol\theta^{(k,k')}$ are then updated with a gradient ascent procedure as:  
	\begin{equation}
	\begin{split} 
		& \boldsymbol\theta^{(k,k')}_{i+1} =  \boldsymbol\theta^{(k,k')}_{i} + \\ 
		& \alpha^{(k,k')}\int dr r [ g^{(k,k')}(r|\boldsymbol\theta^{(k,k')}_i) - g^{(k,k')}_{\text{tgt}}(r)][ \boldsymbol\nabla_{\boldsymbol\theta} 
		u^{(k,k')}(r|\boldsymbol\theta) ]_{ \boldsymbol \theta \equiv \boldsymbol \theta^{(k,k')}_i} \\ 
	\label{eq:rel_entropy_binary}
	\end{split} 
	\end{equation}
where $i$ indicates iteration step, $g(r|\boldsymbol\theta^{(k,k')}_i)$ represents the simulated radial distribution function (RDF), $g^{(k,k')}_{\text{tgt}}(r)$ is the target RDF, $[ \boldsymbol\nabla_{\boldsymbol\theta} u(r|\boldsymbol\theta) ]_{ \boldsymbol \theta \equiv \boldsymbol \theta^{(k,k')}_i}$ denotes the gradient of the pair potential, and $\alpha^{(k,k')}$ is a learning rate chosen to ensure simulation stability and convergence. While it may be possible to choose a single value of the learning rate parameter that is effective for all interaction types, we adopt independently tuned values for the three interaction types here to attain faster convergence. We define $u(r|\boldsymbol\theta^{(k,k')})$ as a set of Akima splines whose knots are computed from $\boldsymbol\theta^{(k,k')}$ as reported in Ref.~\cite{RelEntropy_2D_structures}. Lastly, the knot amplitudes are restricted to increase with decreasing values of $r$ to ensure a monotonically decreasing (i.e., repulsive) pair potential. A derivation of the update expression presented in eq. \ref{eq:rel_entropy_binary} is provided in the supplementary material. 

	Briefly, an RE optimization is carried out as follows. An iteration RE step starts by simulating a high temperature fluid interacting through pair-potentials $u^{(k,k')}(r|\boldsymbol\theta)$ and cooled slowly to a prescribed optimization temperature $T^*$. The system is allowed to equilibrate and rdf statistics are collected. Using the differences between the calculated and target rdfs, potential parameters are updated as per eq. \ref{eq:rel_entropy_binary} and this resulting potential used to simulate the high temperature fluid system where the entire process is repeated in the next iteration step. For implementation details please see subsection \ref{ssec:MS} below. 

\subsection{Crystal Target Selections}
	%---vertices
	\begin{figure*}[htp!]
	\includegraphics[scale=0.28]{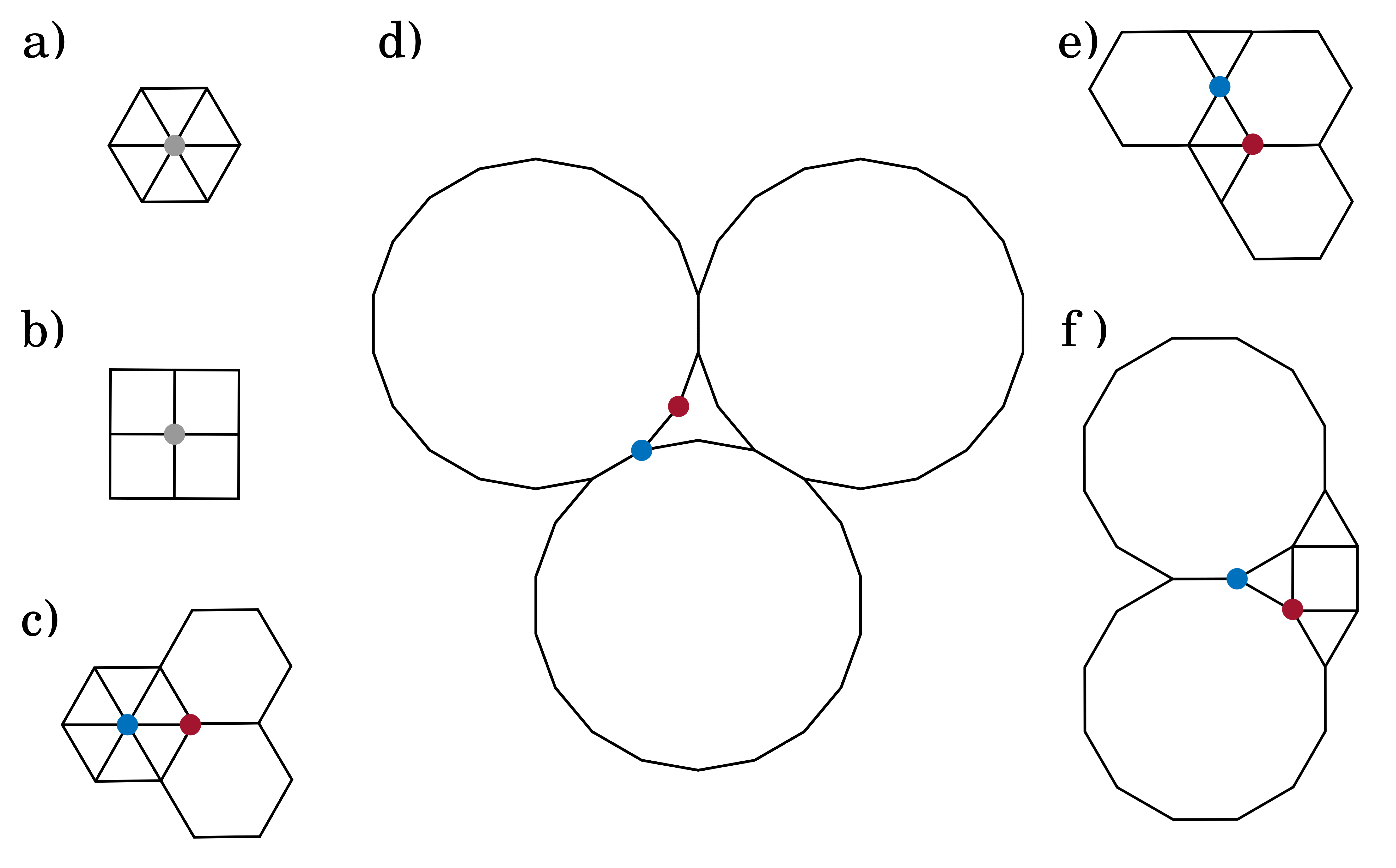} 
	\caption{ 1-uniform (grey dots) showing a) triangular $[3^6]$ and b) square $[4^4]$ vertices. The a) vertex is used to generate the triangular binary crystal while b) can be used to generate square binary, intercalated component rows forming single or double stripes as well as a structure consisting of a large open square with a component in the corners and the other at the sides (dubbed ``square corral'').  2-uniform (blue and red dots) vertices showing c) $[3^6;3^2.6^2]$ dubbed ``triangular honeycomb'' due to internal (blue) triangular and surrounding (red) hexagonal shapes d)  $[18^2.3_{2\pi/9}]$ or ``octadecagonal star binary'' due to the octadecagonal and star polygons motifs e) $[3.6.3.6; 3^2.6^2]$ or ``rectangular kagome'' and f) $[3.12.12; 3.4.3.12]$  dubbed ``square truncated hexagonal'' due to the square super-orientation of the dodecagon shape in the tasselated structure. }
	\label{fig:vertices}
	\end{figure*} 
 A tiling can be described by the vertices formed by the underlying polygon tiles and denoted as [$n_1.n_2$....] where $n_i$ denotes the numbers of sides for the polygons that meet at each vertex.\cite{TilingBook} For instance, in the case of a square lattice, each vertex is the meeting point of 4 squares tiles; in vertex notation this lattice is denoted by [4.4.4.4] or [$4^4$] for short. Tilings consisting of $k$ vertex types are said to be $k$-uniform and denoted similarly as [$n_1^1.n_2^1$...; ...; $n_1^k.n_2^k...$]. For a crystal composed of particles at the vertices of regular polygon tilings, the number of non-equivalent origins in the crystal then corresponds to the number of $k$ vertices necessary to create the crystal. In this work, we only consider target lattices that are 1- and 2-uniform tilings in order to guarantee a single equivalent origin for each component in the binary mixture. In addition to regular polygon tilings, we consider a tiling consisting of an octadecagonal polygon and a star polygon with 3 corners with internal angle of $2\pi/9$--denoted as $[18^2.3_{2\pi/9}]$. Our chosen targets are shown in Fig.~\ref{fig:vertices} a) through f) in formal vertex notation. Note that some realized crystals are based on a single underlying vertex (e.g., square [$4^4$]) but with the components arranged so as to create new structural motifs (e.g., intercalated rows of stripes) while still conserving origin equivalency. In such a case the actual vertex seen by the {\em individual} component may not be regular, but together with the other component, still span the original vertex of the tiling.

\subsection{Molecular Simulations} 
\label{ssec:MS}
The RE optimization algorithm described above can be interfaced with any standard simulation engine, and here we use the GROMACS 5.0.6 molecular dynamics package \cite{GROMACS1,GROMACS2}. We define the component masses as $m=m^{(A)}=m^{(B)}=1$ and nearest same-component neighbor distance as $\sigma=\sigma^{(k,k)}=1$. The energy scale is defined by $\beta=(k_{\text{B}} T)^{-1}$, where $k_{\text{B}}$ is the Boltzmann constant, and we take the effective temperature $T^*=(\beta)^{-1}$ to be unity at the optimization temperature. Briefly, a binary system of particles of type A and B interacting via independent self (AA and BB) and cross (AB) monotonic repulsive spline pair potentials is simulated in the canonical ensemble using a periodically replicated rectangular simulation cell with the aspect ratio chosen to accommodate the target lattice. The corresponding simulation time step is set to $dt=0.001$ for all runs. The number of A and B particles, $N^{(A)}$ and $N^{(B)}$ respectively, is chosen to match the ideal target configuration stoichiometry $N^{(A)}:N^{(B)}$ and kept to a combined total particle number of approximately $N=N^{(A)}+N^{(B)}=1000$ for all targets. The system is initiated at a high temperature ($T_h^*$) and annealed to $T^*=1$ over a span of $5\times10^6$ times steps. For triangular binary or similar compact structures, $T_h^*=3$, while for more open or square-based structures, $T_h^*=1.5$ was sufficient to ensure melting.  Radial distribution function $g^{(k,k')}(r)$ statistics are collected for an additional $10^6$ time steps after cooling to $T^*=1$. Target rdfs are generated by tethering particles to ideal crystal positions using a quadratic restoring potential with spring constants chosen such that peaks are sharp but integrable.\cite{RelEntropy_2D_structures,RelEntropy_general} Appropriate spring constants were generally in the range of 10-1000; larger values were required for the more open targets. Using the simulated structural data, the spline potential was updated per Eq.~\ref{eq:rel_entropy_binary}, and the process was iterated in this manner. In practice, $\alpha^{(k,k')}=0.005-0.015$ is sufficient for all crystal target structures considered here. Convergence is typically achieved in about 80-150 iterations, and optimization is considered complete once the self-assembled crystal remains stable up to the high temperature limit $T_h^*$. For a description of the scheme used to determine interaction cut off radius in our systems, see the supplementary material.  

%------------Results and Discussion Section-----------------
\section{Results and Discussion}
\label{sec:Results} 
	%---hex based binary
	\begin{figure*}[htp]
	\includegraphics[scale=0.28]{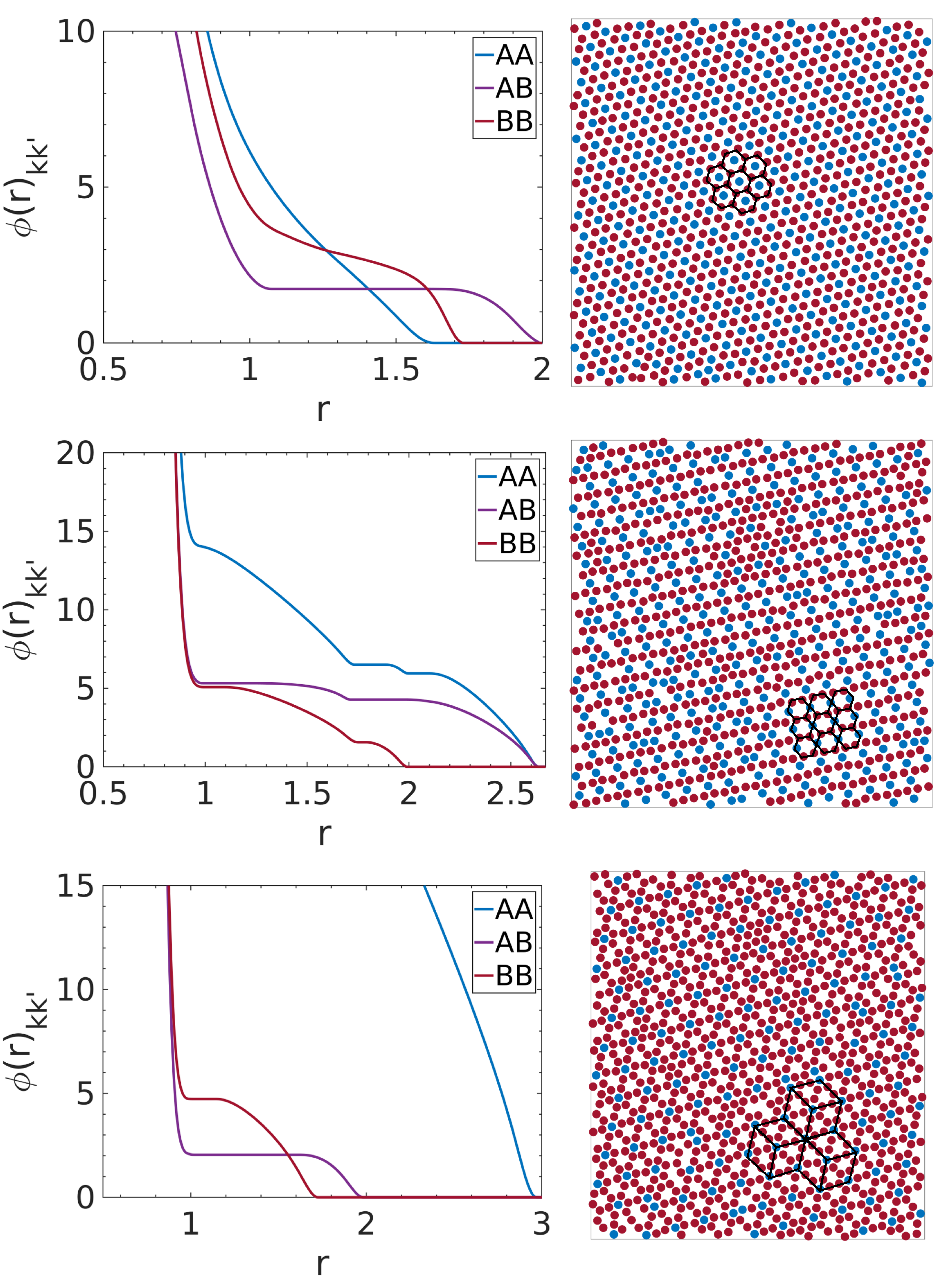} 
	\caption{Optimized pair potentials and representative particle configurations for the triangular binary ($[3^6]$-top),  rectangular kagome ($[3.6.3.6; 3^2.6^2]$-middle), and "triangular honeycomb" ($[3^6;3^2.6^2]$-bottom) lattice assemblies. Black lines are drawn to highlight the ideal crystal structures.}
	\label{fig:hex_binaries}
	\end{figure*} 
	%---square based binary
	\begin{figure*}[htp]
	\includegraphics[scale=0.28]{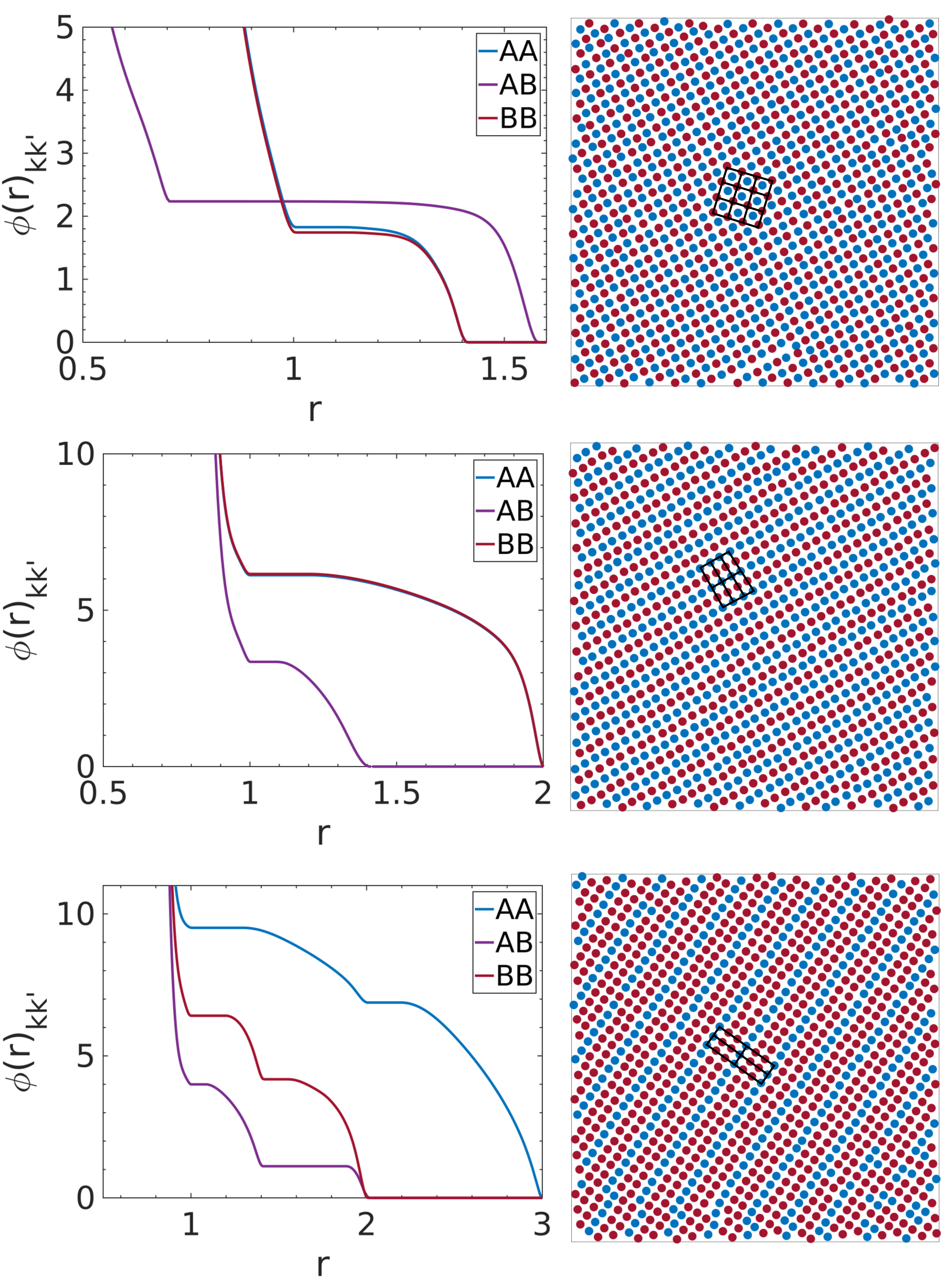}
	\caption{Optimized pair potentials and representative particle configurations for the square binary (top), square single stripe (middle), and square double stripe (bottom) lattice assemblies. Black lines are drawn to highlight the ideal crystal structures.}
	\label{fig:squ_binaries}
	\end{figure*} 
	%---other binary
	\begin{figure*}[htp]
	\includegraphics[scale=0.28]{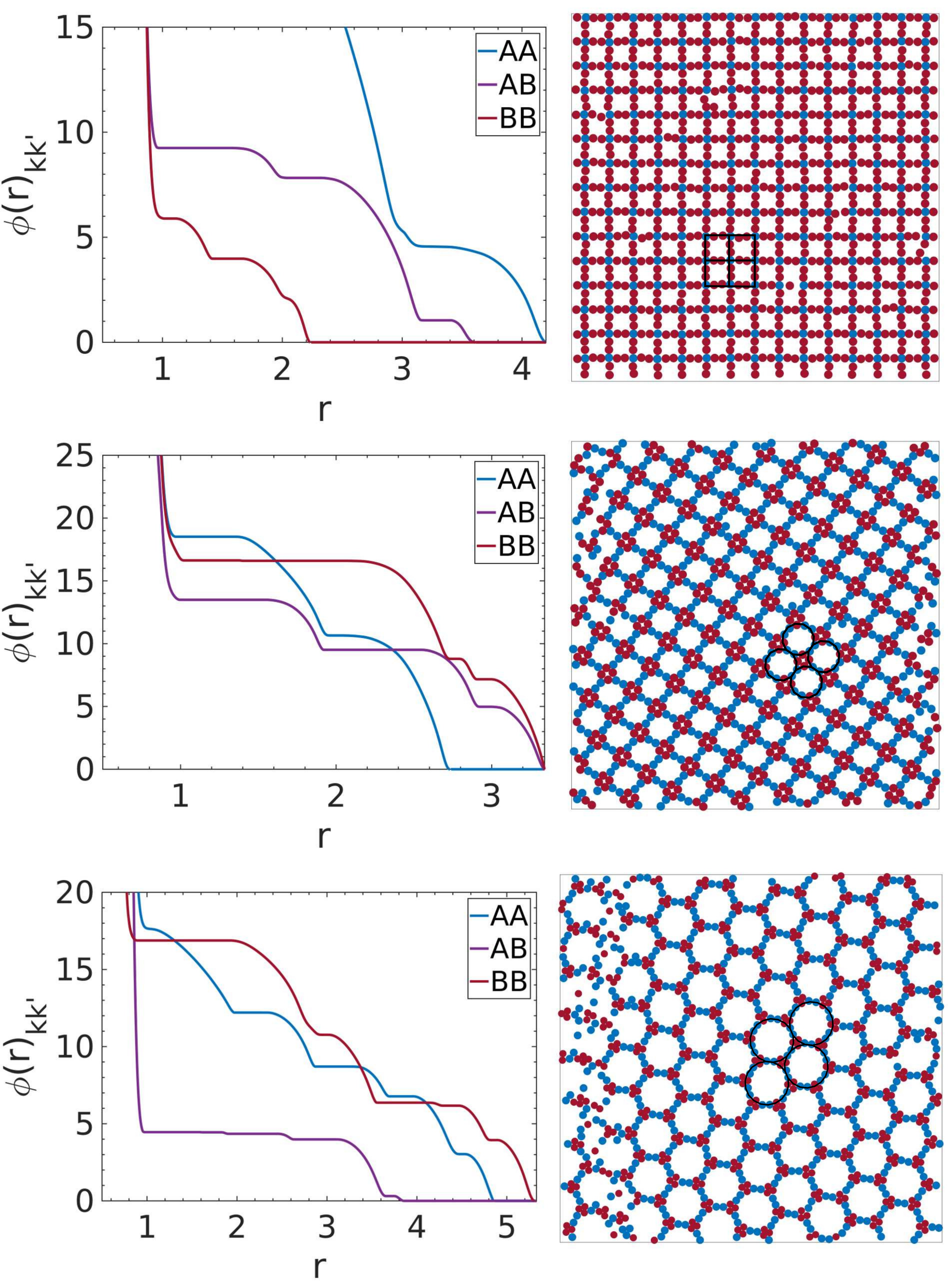} 
	\caption{Optimized pair potentials and representative particle configurations for the square corral (top), square truncated hexagonal ($[3.12.12; 3.4.3.12]$-middle), and octadecagonal star binary ($18^2.3_{2\pi/9}$ - bottom) lattice assemblies. Black lines are drawn to highlight the ideal crystal structures.}
	\label{fig:other_binaries}
	\end{figure*}

	Following the RE optimization protocol elaborated in Sect.~\ref{sec:Methods}, isotropic interparticle potentials for binary mixtures that successfully self-assembled each of our nine binary crystal targets were designed. The optimized potentials and assemblies are shown for systems featuring triangular and honeycomb motifs (Fig.~\ref{fig:hex_binaries}), squares and stripes (Fig.~\ref{fig:squ_binaries}), and other more complex, open structures (Fig.~\ref{fig:other_binaries}). Most of the optimized potentials yield excellent particle assembly including the very open and intricate `square truncated hexagonal' (STH) or `octadecagonal star binary' lattices presented in Fig.~\ref{fig:other_binaries}. 
	
	Considering the designed interactions and the associated assembled structures shown in Fig.~\ref{fig:hex_binaries}--\ref{fig:other_binaries}, two general observations can be made. First, targets featuring equivalent component sites yield comparable interactions for A and B components as should be expected based on symmetry. This can be seen, for instance, in the square binary or square stripe structures shown in Fig.~\ref{fig:squ_binaries} (top or middle, respectively), where exchanging component identities would yield near-identical configurations. Second, optimized AA, BB and AB pair potentials become longer ranged and exhibit more features (i.e., shoulders and plateaus) as the target structures become more open and complex. Compare, for instance, potentials for square binary in Fig.~\ref{fig:squ_binaries} (top) with those of STH in Fig.~\ref{fig:other_binaries} (middle). The former are clearly short ranged and feature a single shoulder, while STH interactions are longer ranged and exhibit multiple plateaus. 
	
	We also find evidence for two advantages of binary mixtures over single-component systems for self assembly: 1) the optimized pair interactions of a binary system can each be simpler than an analogous single-component interaction designed to stabilize the same \emph{global} (or overall) structure, and 2) the expanded parameter space of a binary system can help to stabilize a richer variety of self-assembled structures than single-component systems. To illustrate the former, we consider the triangular honeycomb structure in Fig.~\ref{fig:hex_binaries} (bottom). The underlying tiling is 2-uniform as illustrated in Fig.~\ref{fig:vertices}c. Therefore, the global structure of the lattice can be naturally partitioned into two simpler sub-lattices: a honeycomb lattice (red particles) and a triangular lattice with a side length of three (blue particles). By allowing two components to occupy the two distinct types of lattice sites, relatively simple interactions can be combined to favor self-assembly of a rather complex target. By contrast, when the same \emph{global} structure was targeted via a single-component optimization, the best resulting interaction was not only considerably more complex, but also self-assembly of the target structure was significantly less satisfactory than for the binary mixture (see Fig.~S1). As a second example, a previous study reported\cite{ZT_MuOptKagomeAsymLats} that single-component assembly of a rectangular kagome lattice from a single-component system required an interaction possessing an abrupt attractive well within a larger repulsive profile, while the optimized binary interactions reported here for the same structure are purely repulsive and limited to a few shoulder features; see Fig.~\ref{fig:hex_binaries} (middle).

	With respect to binary mixtures increasing the diversity of possible self-assembled structures relative to single-component systems, the partitioning of a desired global structure into individual sub-lattices corresponding to each species not only allows for self-assembly of intricate structures such as STH and octadecagonal star binary lattices (Fig.~\ref{fig:other_binaries}), but it also opens up the possibility of segregating components in specific \emph{local} structures using an otherwise simple global lattice structure. This is seen clearly in Fig.~\ref{fig:squ_binaries} where the same underlying square lattice is partitioned to form intercalated squares (top) or stripes with either a 1:1 (middle) or a 2:1 (bottom) ratio. As expected, we find that increasing the asymmetry of the particle arrangement (as is displayed from top to bottom in Fig.~\ref{fig:squ_binaries}) generally requires more complex potentials in terms of both the interaction range and the number of shoulders. A similar argument applies to the triangular binary target, where it is known that a single-component hard-core fluid favors a \emph{global} triangular lattice, but achieving assemblies possessing the specific relative AA and BB ordering characteristic of the target lattice requires the binary interactions shown in Fig.~\ref{fig:hex_binaries}(top). In short, the above observations indicate that the enhanced design space of binary systems can enable the self-assembly of structures also available to single-component systems (but with {\emph{simpler}} interactions) as well as the self-assembly of significantly {\emph{more complex}} structures than those attainable in a single-component system.
	
%-------------- 
	%---single component  plot comparison 
	\begin{figure*}[ht]
	\includegraphics[scale=0.80]{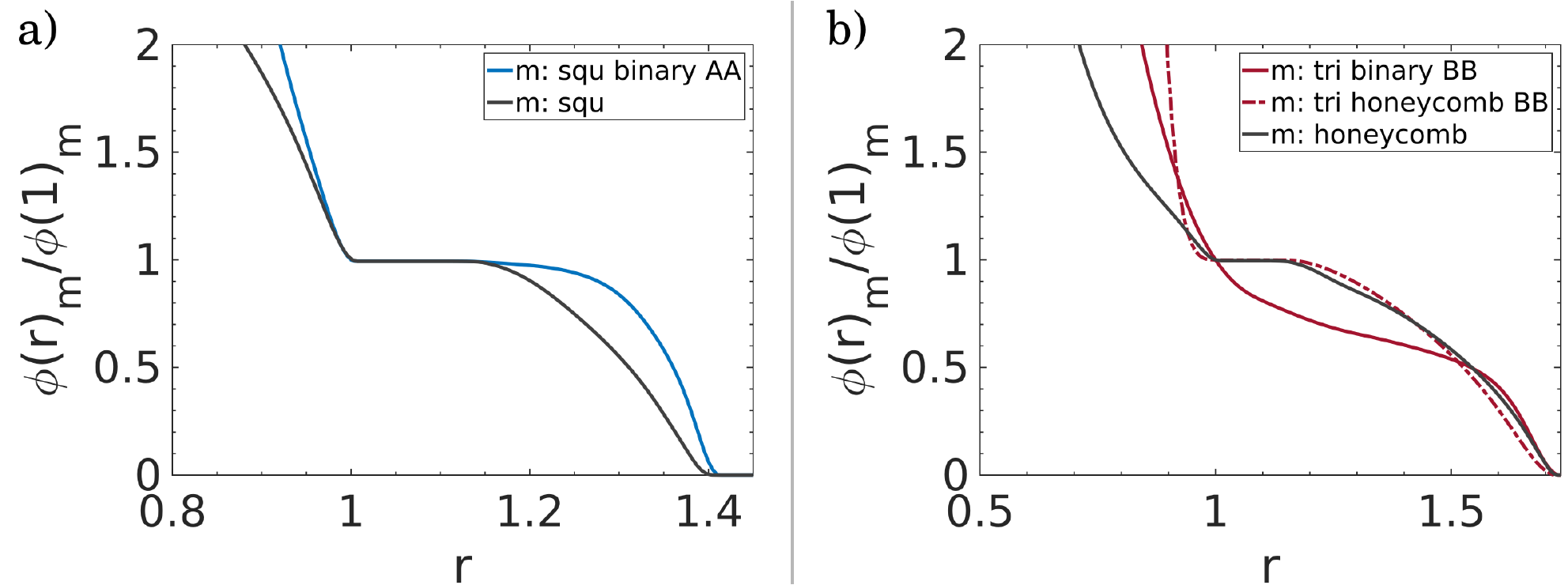} 
	\caption{a) AA component of the optimized pair interaction for the square binary structure (blue) compared to that reported\cite{RelEntropy_2D_structures} for the single-component square structure (black). b) BB component of the optimized pair potential for the triangular binary (solid red) and triangular honeycomb binary (dash red) lattices compared to that reported for the honeycomb potential\cite{RelEntropy_2D_structures} (black).}
	\label{fig:single_component_pots}
	\end{figure*}

	%---single component assembly comparison 
	\begin{figure*}[htp]
	\includegraphics[scale=0.90]{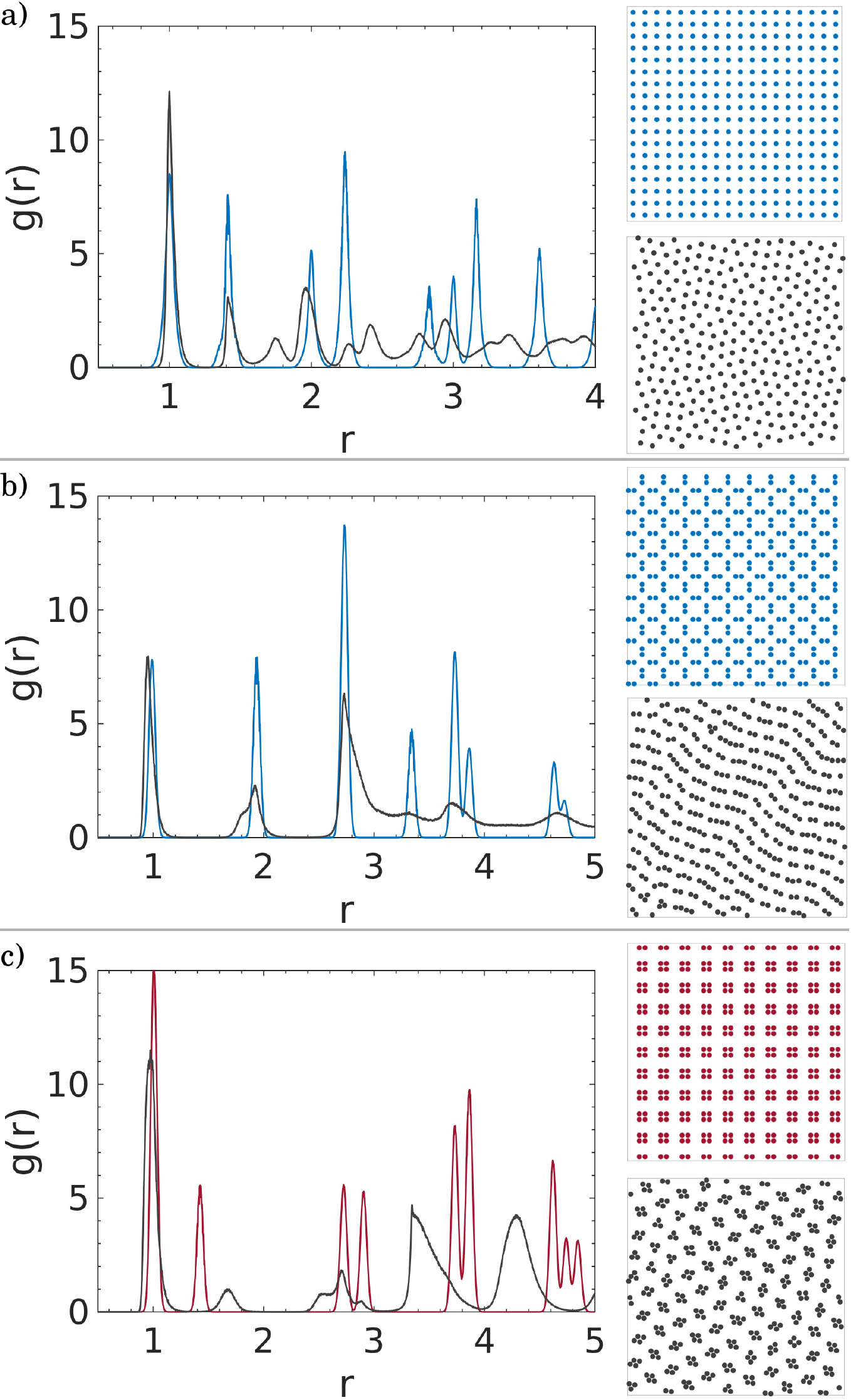} 
	\caption{Radial distribution functions and configurations of single-component assemblies of particles interacting via optimized binary self interactions (blue or red) versus behavior in the fully optimized binary system (grey) for the a) square binary AA interaction (same as BB by symmetry) b) square truncated hexagonal AA interaction and c) square truncated hexagonal BB interaction.}
	\label{fig:single_component}
	\end{figure*} 
	
	%---no AA,AB etc square 
	\begin{figure*}[htp]
	\centering
	\includegraphics[scale=1.10]{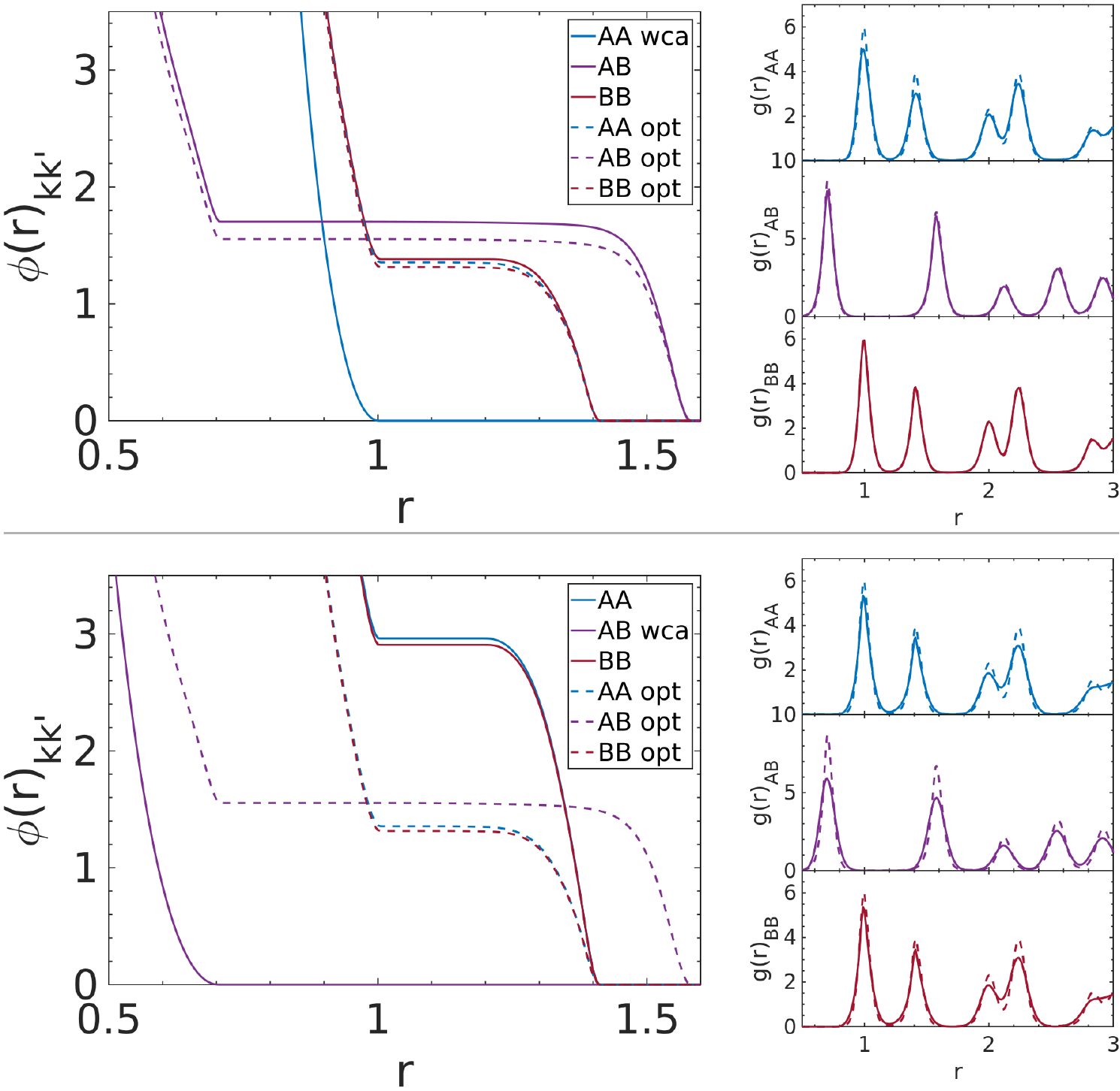}
	\caption{Optimized component interaction and radial distribution function comparison for square binary optimizations where the AA (top) and  AB (middle) interactions have been fixed to display a simple WCA-like repulsive form described in the text. Note that, for both cases, the square binary assembly with fully optimized interactions leads to sharper RDF peaks at the target temperature.} 
	\label{fig:squ_no_AA_etc}
	\end{figure*} 

	Focusing on binary assemblies, we can also gain some insight into the mechanisms for their global stabilization by comparing the underlying component interactions to those of single-component systems optimized to stabilize similar local structures. Specifically, equation \ref{eq:rel_entropy_binary} recognizes that the binary system can comprise independent self and cross interactions that are optimized to recreate the RDFs of the target lattice. However, the equation alone does not clarify the extent to which the binary assembly can be considered a trivial superposition of sub-structures that can be stabilized by component self interactions (e.g., approximately equal to those expected from the analogous single-component target structure) versus a more cooperative, or coupled, assembly relying on nontrivial cross interactions. In order to help address this question, we compare the optimized self interactions of the binary system to analagous single-component interactions that stabilize the same \emph{local} structure. 

	We begin by considering the square binary, triangular binary, and triangular honeycomb targets shown in Fig.~\ref{fig:squ_binaries} (top), \ref{fig:hex_binaries} (top) and \ref{fig:hex_binaries} (bottom), respectively. The underlying single-component structure is that of a square lattice for square binary AA or BB components and a honeycomb lattice for the BB component of both triangular binary and triangular honeycomb structures. As such, we can plot these individual component interactions and compare them to single-component interactions known to stabilize these lattices.~\cite{RelEntropy_2D_structures} This comparison is shown in Fig.~\ref{fig:single_component_pots} for the optimized square lattice (a) and the honeycomb lattice (b) interactions where potentials are normalized such that $\phi(r)/\phi(1)=1$. As seen, the square binary AA(BB) interaction is largely similar, though not identical, to its single-component equivalent in range and complexity. The BB interactions for the triangular honeycomb and triangular binary (Fig.~\ref{fig:single_component_pots} b)) also approximate those of the reported honeycomb potential, though the triangular binary BB interaction deviates more strongly from the single-component results. This larger discrepancy, as we will show below, is an indication that individual component interactions need not exactly match their ideal local target lattice counterpart in order to achieve proper global assembly. 

	 To investigate the above deviation more closely, we use one of the self (AA or BB) interactions from the optimized binary system to carry out a single-component assembly simulation at the same temperature and box size as the binary system, where the other component has been removed from the simulation box. We then compare the resulting equilibrium assemblies to the expected perfect local lattice for that component. To this end, we choose the simpler square binary structure (Fig.~\ref{fig:squ_binaries} top) and the more elaborate STH structure (Fig.~\ref{fig:other_binaries} bottom) as two contrasting test cases. As seen in Fig.~\ref{fig:single_component}a) (top right), the square lattice is the local structure for the A(B) component in the square binary target. However, as shown in the bottom right of a), particles interacting via the AA interaction form a largely amorphous configuration. Despite this, the corresponding RDF for the extracted binary interaction (blue) shows a good match between the first two target square lattice coordination shells positions (note that, for this case, the AA pair interaction only spans the first two target coordination shells). 

    Similarly for the STH target structure, the A component locally forms a stretched truncated square lattice. However, particles interacting with the AA potential in a single-component simulation assemble into stripe-like phases. Nonetheless, comparing RDF peak positions shows good agreement with the target in the first three coordination shells which spans the full interaction range. Lastly, looking at the STH BB component assembly in c), we expect square clusters in a square super-lattice arrangement, but simulations yield rhomboid clusters in a triangular super-lattice orientation instead. While the peak positions in the RDFs do not overlap as closely as in Fig.~\ref{fig:single_component}a,b, visual comparison of the target and the formed assembly shows that the mean inter-cluster distance is approximately the same in both cases. Together, these results demonstrate that individual component interactions cannot be expected to fully recreate the underlying local target structure in the binary target but only to ensure proper local particle positioning on average. This is why the optimized interactions need not (and generally will not) closely match corresponding single-component target interactions; the latter fail to fully stabilize the local target structures and more specific coupling between interaction types must play a key role. 

    To further unravel the requirements for assembly of binary structures, we performed optimizations where one of the three interaction types was fixed as a hard-core-like [Weeks-Chandler-Andersen (WCA)] potential, while the rest were optimized as usual.\cite{WCA-note} In particular, we consider the square binary target from Fig.~\ref{fig:squ_binaries} (top) and carry out optimizations controlled for annealing schedule and iteration step so as to isolate the effects of fixing the interaction relative to the fully optimized system. Results for the square binary structure, where either AA(BB) or AB interactions are fixed, are shown in Fig.~\ref{fig:squ_no_AA_etc} (top or bottom, respectively). In the first case, fixed AA interactions resulted in stronger AB and BB interactions that helped boost global structure stability--though clearly not as efficiently as the fully optimized system (compare AA RDF, top right). This contrast is more drastic when AB (the cross-coupling) is fixed, resulting in sharpened AA and BB interactions that nevertheless fail to restore original system stability (broadened peaks for all RDFs, bottom right). Fixing one of the individual interactions as hard-core-like and optimizing the others does not work at all in assembling in the more complex binary structures. For instance, when such a procedure was carried for the STH lattice, it resulted in phase separation for fixed AA or BB interaction, or stripe-like configurations for fixed AB interaction (see Fig.~3S). Similar results held for the square corral structure.   

	Together, these results help to highlight the individual roles of self and cross interactions, respectively, in binary assemblies. For simple global targets like the square binary lattice, stronger AB and BB optimized interactions can compensate for a hard-sphere-like AA interaction to result in a successfully self-assembled target structure.  However, for more intricate local orderings like those in the STH lattice, sharpening remaining interactions is not enough to compensate for a neutral interparticle interaction, resulting in a failed global assembly. Similarly, fixing the cross interaction places a larger load on self-self interactions to achieve correct local positioning, but these cannot stabilize relative component ordering {\emph{by definition}} and as such this strategy results in reduced overall global stability for simple and complex targets alike.

    To summarize, the analysis of Fig.~\ref{fig:single_component}-\ref{fig:squ_no_AA_etc} strongly suggests that global binary assembly can be understood as follows: self interactions act as a 'primer' that help ensure that individual component particles assemble into the right positional order (coordination shells), while cross interactions `bind' the locally ordered particles into their correct, target positions.\cite{Analogy-note} This is why, despite the self interactions failing to assemble the target local structure in an analogous single-component system, they nonetheless encode the relevant length scales present in the local structure. In fact, substituting the optimized single-component square lattice interaction (which necessarily contains the characteristic length scales of a square) for the self-interactions in the intercalated square binary system results in successful assembly of the target, with greatly enhanced local AA(BB) stability as seen in the RDFs in Fig.~S2. The analysis also highlights the importance of the interplay among the self and cross interactions for intricate structures, as it is through mutual coupling that the entire global structure is locked into place. This result also implies that the self interactions that establish a necessary \emph{local} structure may be much simpler than those required for an equivalent single-component system, so long as full system interactions are in place. Indeed, while the stretched truncated square \emph{local} structure in the binary square corral could be achieved by the BB component interaction only spanning four coordination shells, the equivalent single-component truncated square interaction required five shells and resulted in poorer assembly (see Fig.~4S). 
	
%-----------------------------		
\section{Conclusions}
\label{sec:Conclusion}
	In this work, we extended a recently introduced inverse design methodology, relative entropy optimization, to discover new isotropic interactions that favor the formation of targeted multicomponent phases. We have used this approach here to determine interactions for binary mixtures that stabilize a wide variety of two-dimensional lattice assemblies and, in doing so, have gained new insights into how adopting a multicomponent system can affect the overall prospects for self assembly. Although the expanded parameter space for design in multicomponent systems increases the complexity of the design problem, it helps discover simpler interparticle interactions (as compared to single-component systems) to assemble a desired target phase. It also allows for designed assembly of complex phases with structures that cannot be stabilized by single-component materials.

	Mechanistically, our results suggest that optimized interactions between like components in a binary mixture act as a `primer' to help ensure that such species adopt, on average, the correct positional order for the target phase. Cross interactions, in turn, act as a `binder' to further ensure that species conform to the precise local compositional order required by the target. For complex or open lattices, independent design of cross and self interactions is required to stabilize the desired assemblies. 

	Physically, reported single shoulder interactions for targets such as figure \ref{fig:hex_binaries} (bottom) or figure \ref{fig:squ_binaries} (top), are similar to those displayed by core-corona\cite{square_step_stripes} or dendritic particle systems\cite{square_step_quasicrystals}, and predict structure stability at (osmotic) pressures moderately above one atmosphere, assuming room temperature and particle size scales on the order of $\sim$10 nm. However, similar experimental analogues for the more complex interactions required for the remaining structures in this work are more challenging to envision. In future work, it will be interesting to pursue related inverse design calculations for classes of materials whose self and cross interactions are effectively constrained in ways that can be encoded by fundamental physics or empirical mixing rules. 

%----supplemental info-----
\section{Supplementary Material}
Please see the supplementary material for a derivation of the RE update scheme and a description of how interaction cut offs are determined. Additionally, we include figures for the triangular honeycomb structure assembly from a single component, self-assembly results for the square truncated hexagonal structure with fixed hard-core-like interactions, comparison of single component assembly of the B component sub-lattice in the square corral target, and an example of how using optimized single component interactions may help boost binary assembly stability. 

\begin{acknowledgments}
T.M.T. acknowledges support of the Welch Foundation (F-1696) and the National Science Foundation (DMR-1720595 and CBET-1403768). We also acknowledge the Texas Advanced Computing Center (TACC) at the University of Texas at Austin for providing computing resources used to obtain results presented in this paper.
\end{acknowledgments}
%-----------End of Main Body-------------------

%\bibliography{sources}

\begin{thebibliography}{77}
\bibitem [{WCA()}]{WCA-note}%
  \BibitemOpen
  \href@noop {} {}\bibinfo {note} {{WCA-like here is defined as the normal
  repulsive WCA pair potential but with the Lennard-Jones power exponents
  changed from 12-6 to 4-2 and scaled by a factor of 100 for softer
  repulsions.}}\BibitemShut {Stop}%
\bibitem [{Ana()}]{Analogy-note}%
  \BibitemOpen
  \href@noop {} {}\bibinfo {note} {{There may be some apparent awkwardness in
  the use of a `binder' analogy for purely repulsive systems. However, as noted
  in previous works, for the assemblies considered here, the effective forces
  mutually reinforce and lock particles into the correct position making the
  `binding' analogy more fitting.}}\BibitemShut {Stop}%
\end{thebibliography}
%\includepdf[pages={-}]{multicompo_supp}
%\foreach \x in {1,...,8}
%{%
%\clearpage
%%\includepdf[pages={\x,{}}]{multicompo_supp}
%\includepdf[pages={\x,{}}]{multicompo_supp}
%}


\end{document}